\documentstyle[aps]{revtex}
\def \be {\begin{equation}}
\def \ee {\end{equation}}
\def \ni {\noindent}

\def \bea {\begin{eqnarray}}
\def \eea {\end{eqnarray}}
\begin{document}

\title{Matching Spherical Dust Solutions to Construct Cosmological
Models}

\author{D. R. Matravers\thanks{david.matravers@port.ac.uk.}
and N. P. Humphreys\thanks{humphrn@uk.ibm.com}}

\address{Relativity and Cosmology group, School of Computer Science
and Mathematics, University of Portsmouth, Portsmouth~PO1~2EG,
Britain}

\maketitle

\begin{abstract}
 Conditions for smooth cosmological models are set out and
 applied to inhomogeneous spherically symmetric models constructed by
 matching together different Lema\^{i}tre-Tolman-Bondi solutions to the
 Einstein field equations.  As an illustration the methods are applied
 to a collapsing dust sphere in a curved background.  This describes a
 region which expands and then collapses to form a black hole in an
 Einstein de Sitter background.  We show that in all such models if
 there is no vacuum region then the singularity must go on accreting
 matter for an infinite LTB time.
\end{abstract}

\section{Introduction}

Recently we and others \cite{rib,drm} have had cause to use
 matched spherically symmetric solutions of the field equations to model
 fractal \cite{pee} distributions of matter in cosmological situations.
 Single spherically symmetric models (see \cite{cel} for a recent example)
 and matched ones have long been used in cosmology (see \cite{kr1} for an
 extensive and  comprehensive review).  They are remarkably rich
 in structure but there are many subtleties in their application
 to cosmology \cite{hel,bo1,bn1}.  The purpose of this paper is
 to clarify and unify existing results on the construction of {\em
 smooth} models.  The topic has important implications for the
 modelling of structures like voids or the formation of black
 holes in curved backgrounds.  The new results include matching
 between exact solutions in the Kantowski-Sachs family and the
 Lema\^{i}tre-Tolman-Bondi (LTB) solutions and to the occurrence of
 centres.   It is not always realised
 that a spherically symmetric model need not possess a centre
 \cite{h87}.  Here an extended list of the possible types of
 centre is given.  Also all possible regular composite dust models are
 given and they are classified into four classes according to the
 number of centres they allow.  One
 of the classes in which the spatial sections have the topology
 of a 3-torus appears to be new.

This work complements that by Lake \cite{lak} and by Fayos et al.
\cite{fra}. The work by Lake is more theoretical.  That by Fayos,
Senovilla
 and Torres is more general and very interesting geometrically but we
 find  it less easy to see some subtle problems such as the one we
 illustrate in the  final section.  It has long been accepted that
 a black hole in an Einstein-de Sitter model can be constructed
 using a Schwarzchild interface. What had not been shown but is
 demonstrated here is how a black hole evolves from an expanding
 region in an Einstein de Sitter background.  The model is a
 modification of one produced by Papapetrou \cite{pap} which is
 non-physical because the singularity takes an infinite LTB time to be
 completed.  In section 7 it is shown that this infinite time is
 an inevitable result of the matching conditions if no vacuum region is
 included, a result which appears to be not widely known.

 \section{Spherical Dust Solutions}

In comoving coordinates $x^{a} = \{t, r, \theta, \phi \}$ the
 spherically symmetric metric can be written \cite{bo1}
 \be
 ds^{2} = - dt^{2} + X^{2}(r,t) dr^{2} + R^{2}(r,t) d \Omega^{2},
 \ee
 and the dust 4-velocity as $u^{a} = \delta_{t}^{a}$.  Here
 $d \Omega^{2} \equiv d \theta^{2} + \sin^{2} \theta \; d \phi^{2} $
 and $R \geq 0$ with $R = 0$ only at a centre. The coordinate ranges are;
 $r \geq 0$, $0 \leq \theta \leq \pi$, $0 \leq \phi \leq 2 \pi$ and
 $t > T(r)$ (see the solutions below for a definition of $T$).  The dust
 energy momentum tensor is $T^{a}_{b} = \mu u^{a}u_{b}$, where $\mu$
 is the proper matter energy density.

In order for the Einstein tensor to be well defined we require
that
\begin{quote}
$X \neq 0 \neq R$; $R$ is $C^{2}$ in $r$ and $t$; $X$ is $C^{2}$ in
$t$ and $C^{1}$ in $r$.
\end{quote}
We call such behaviour regular.  Regions of regularity may be
joined together to form composite space-times in which the
differentiability conditions hold only piece-wise. More details
are given below when we discuss the matching.  To set the notation
and for later use we first write down the first integrals and
write out the solutions \cite{le1,to1,bo1}.  We will call the
solutions LTB models and refer to the coordinates as LTB
coordinates.  We use units such that $G = c = 1$ and the notation,
overdot for $u^{a}\partial_{a} = \partial/\partial t$ and prime
for $\partial/\partial r$.  If $R^{\prime} \neq 0$, then the
Einstein field equations reduce to
\be
X = \frac{R^{\prime}}{\sqrt{1 + E(r)}},  \label{fe2}
 \ee
  where $E(r)$ is an arbitrary function of integration, and
 \be
 \dot{R}^{2} = \frac{2 M(r)}{R} + E,   \label{fe1}
 \ee
 with $M(r)$ a second function of integration.  The proper density is
 given by
 \be
 \mu = \frac{M^{\prime}}{4 \pi R^{\prime}R^{2}}.  \label{den1}
 \ee
The equations (\ref{fe2}) - (\ref{den1}) have five solutions:
\\

\ni {\bf [s1]} for $\{E = M = 0\}, $ \hspace{6ex}  $ R = - T(r)$,
\\

\ni {\bf [s2]} for $\{E > 0, M = 0\}, $ \hspace{5ex} $ R =
\sqrt{E}(\epsilon t - T)$, \\

\ni {\bf [s3]}  for $ \{E = 0, M > 0 \}, $ \hspace{5ex}
 $ R = (9M/2)^{1/3}(\epsilon t - T)^{2/3}$, \\

\begin{tabbing}
\ni {\bf [s4]} for $\{E > 0, M > 0 \}, $ \hspace{5ex}
 \= $ R  =  \frac{M}{E}(\cosh \eta - 1)$, \\
 \> $\sinh \eta - \eta  =  (\epsilon t - T) E^{3/2} M^{-1}$, \\
 \> $0 < \eta < \infty $, \\
 \end{tabbing}

\begin{tabbing}
\ni {\bf [s5]} for $ \{E < 0, M > 0 \}, $ \hspace{5ex}
 \= $R  =  M(\cos \eta - 1) E^{-1}$, \\
 \> $\eta - \sin \eta =  (\epsilon t - T) |E|^{3/2}M^{-1}$, \\
 \>  $0 < \eta < 2 \pi $.\\
\end{tabbing}

\ni Here, $T(r)$ is a further function of integration and
$\epsilon = \pm 1$.  The solutions {\bf s1} and {\bf s2} are
locally Minkowskian. The surfaces $ \{\epsilon t - T = 0\}$ are
spacelike and singular. Following Bondi \cite{bo1}, $M$ is the
relativistic generalisation of Newtonian mass and $\frac{1}{2}E$
is the total energy.

The case $R^{\prime} = 0$ leads to an inhomogeneous generalisation
of the Kantowski-Sachs metric \cite{el1} \cite{hu1} which does not
have centres of symmetry in the hypersurfaces $t = $ constant. It
is given in our notation by
\\

 \ni  {\bf [s6]}
 \bea R & = & \bar{M} (1 - \cos \eta),
\\ \eta - \sin \eta & = & \bar{M}^{-1}(\epsilon t - T),
\\ X & = & A(r) \frac{\sin \eta}{1 - \cos \eta} + B(r)\left[1 -
\frac{\eta \sin \eta}{2(1 - \cos \eta)} \right],
 \eea
  where $A$
and $B$ are integration functions, $\bar{M}$ is a constant, and $0
< \eta < 2 \pi$.  We label this solution {\bf s6}.
\section{Junction Conditions}

Our aim is to establish a class of cosmological models which do
not have shell-crossing or surface density layers, but which can
be constructed by matching together different LTB solutions from
the set {\bf s1} to {\bf s6}, so we impose the Darmois conditions
\cite{bo2} on the junctions.  For a comoving space-like junction
$r = $ constant and dust, these require that the first and second
fundamental forms are continuous across the junction (interface).
The unit normal is given by $n_{a} = |X|\delta^{r}_{a}$, and the
first fundamental form is the metric intrinsic to the interface,
i.e.,
 \bea
{h}_{ab} & = & g_{ab} - n_{a}n_{b} \\ & = & {\rm diag}(-1, 0,
R^{2}, R^{2} \sin^{2} \theta),
 \eea
 and the second fundamental form is the extrinsic curvature which
 is given by
 \bea
 {K}_{ab} & = & {h}^{c}_{a} {h}^{d}_{b} \nabla_{d}
n_{c} \\
 & = & {\rm diag}\left(0, 0, \frac{RR^{\prime}}{|X|}, \frac{RR^{\prime}}
 {|X|}\sin^{2} \theta \right).
 \eea
From these it follows that the necessary and sufficient conditions
for matching are that
\be
R \hspace{5ex} {\rm is \; continuous \; in } \hspace{1ex} r,
\label{cond1} \ee
\be
\frac{R^{\prime}}{|X|} \hspace{5ex} {\rm is \; continuous \;in}
\hspace{1ex} r.  \label{cond2} \ee

\ni The nature of the problem changes in the non-comoving case
\cite{hu1} \cite{is1}.  The conditions at the junction must be
satisfied through some range of values of $r$.  Explicitly, if the
spacelike non-comoving boundary surface is given by
 \be
 r - g(t) = \hspace{1ex} {\rm constant},   \label{surf}
 \ee
 where for convenience we choose $P$ such that
 \be
 \frac{d g}{dt} = \frac{P}{|X|\sqrt{P^{2} - 1}},
 \ee
 then the unit normal to the surface is given by
  \be
  n^{a} = \left(P, \frac{1}{|X| \sqrt{P^{2} - 1}}, 0, 0 \right)
  \ee
  and $n^{a}n_{a} = -1$.  The unit tangent to the surface with
  $\theta = \phi = $ constant is
  \be
  m^{a} = \left(\sqrt{P^{2} - 1}, \frac{P}{|X|}, 0, 0 \right).
  \ee
  The intrinsic curvature of the surface is given by
  \be
  \hat{h}_{ab} = \left(\begin{array}{cccc}
                        P^{2} - 1 & - P|X| \sqrt{P^{2} - 1} & 0 & 0 \\
                        - P|X|\sqrt{P^{2} - 1} & P^{2}X^{2} & 0 & 0 \\
                        0 & 0 & R^{2} & R^{2} \sin^{2} \theta
                        \end{array} \right),
  \ee
  and the extrinsic curvature is
   \be
   \hat{K}_{ab} = \left( \begin{array}{cccc}
                        (P^{2} - 1)F_{1} & - F_{1} |X| P \sqrt{P^{2} -
                        1}& 0 & 0 \\
                        - F_{1} |X| P \sqrt{P^{2} - 1} &
                        P^{2}X^{2} F_{1} & 0 & 0 \\
                        0 & 0 &  F_{2} & 0 \\
                        0 & 0 & 0 & F_{2} \sin^{2} \theta
    \end{array} \right),
   \ee
   where $ F_{1} = \frac{dP}{dt} + P \frac{\dot{X}}{X}$ and $F_{2} =
   PR \frac{dR}{dt} - \frac{R R^{\prime}}{|X| \sqrt{P^{2} - 1}}$,
   and all the ordinary derivatives are taken along the paths
   $\{\theta, \phi \} = $ constant in the tangent space to the
   hypersurface.  On the hypersurface, one coordinate is surplus
   because of (\ref{surf}).

We will not deal with non-comoving junctions further here except
to mention two points.  First that at the junction at least one of
$E, M$ and $T$ must be constant for a range values of $r$.  This
follows from the fact that they have to be continuous by the
Darmois conditions and at least one of them has at most one value
in common (or as a limit for {\bf s6}) for allowed\footnote{see
later} matches  between  the solutions {\bf s1} to {\bf s6}.  Thus
in the non-comoving case there will be an interface region.
Second, that Krasi\'nski \cite{kr2} suggests that non-comoving
boundaries could find an application in the formation of
structure. The idea is that it may be possible to use them to
allow incoming matter to augment the condensations discussed by
Bonnor \cite{bo3}.

From here on we restrict attention to junctions where there are no
surface layers or interface region and so they are necessarily
comoving and the metrics on both sides are of the junction are
determined by the integration functions $E(r)$, $M(r)$ and $T(r)$
(and $A(r)$ and $ B(r)$ in regions where $E(r) = -1$).  For the
spherically symmetric dust models the junction conditions
(\ref{cond1}) imply that
\be
M(r)(\geq 0),\; E(r)(\geq - 1) \ \mbox{\rm{and}} \ T(r) \
\mbox{\rm{are continuous}}.
 \ee

\section{Regularity Requirements}

In this section the physical requirements to be imposed on the
metrics are made more explicit and justified.  It follows from the
conditions on the metric and the matching conditions across
spacelike surfaces that the metrics {\bf s1} to {\bf s6} are at
least $C^{2}$ in $t$. Thus we are only concerned with the
behaviour with respect to the radial coordinate $r$.  We start by
requiring that lim$_{\pm}X \neq 0$ everywhere including across the
junctions. This condition is important because it enables us to
express the continuity properties of physical quantities
unambiguously through their differentiability in $r$, and prevents
shell crossing.  To ensure that there is not a curvature
singularity as $R \rightarrow 0$, we require that $M/R^{3}$ is
finite everywhere except trivially at the spacelike singularity
$\tau \rightarrow 0$. Finally we require that at a centre the
shear
  $$
  \sigma^{b}_{a} \equiv \frac{1}{3}\left[\frac{\dot{X}}{X} -
  \frac{\dot{R}}{R} \right] \; \times \; \mbox{diag}(0, 2, -1, -1)
  $$
 go to zero to maintain spherical symmetry. If it is
 not zero, then the eigendirections of the shear tensor will violate
 spherical symmetry at the centre. \\

In summary, we require the following properties to hold in our
matched spacetimes:
\begin{description}
  \item[R1] The junction conditions (\ref{cond1},\ref{cond2}) hold.
  \item[R2] lim$_{\pm}X \neq 0$.
  \item[R3] The shear tends to zero whenever $R \rightarrow 0$, i.e. at a centre.
  \item[R4] $M/R^{3}$ remains finite as $R \rightarrow 0$ (except trivially at the
  spacelike singularity $\tau \rightarrow 0$).
  \item[R5] The metrics {\bf s1} to {\bf s6} are regular within their domains,
   i.e. between junctions.
\end{description}

We will now describe the implications of these conditions within
the domains of metrics. It follows from the field equation
(\ref{fe2}) that
 \be
R^{\prime} \hspace{2ex} \mbox{\rm{can change sign only at values
of}} \ r \ \mbox{\rm{for which}} \  E(r)= -1, \label{15}
 \ee
 i.e., in {\bf s5}. For {\bf s1} to {\bf s4}, $R^{\prime} \geq 0$ or
$R^{\prime} \leq 0$ throughout.  For metrics {\bf s1} to {\bf s4},
differentiation of the exact solutions for $R$, and examination of
the asymptotic behaviour for large and small $t$, yield
\be
\pm R^{\prime} > 0 \Rightarrow \hspace{2ex} \{ \pm M^{\prime} > 0,
\hspace{3ex} \pm E^{\prime} > 0, \hspace{3ex} \pm T^{\prime} < 0
\}.  \label{asy}
 \ee
 For {\bf s5} in the region in which $E(r) \neq -1$, i.e. where
$R^{\prime}$ does not change sign, differentiation of the exact
solution for $R$ with respect to $r$ and investigation of the
result as $\eta \rightarrow 2 \pi$, and assuming $\mu \geq 0$,
gives
\be
\pm R^{\prime} \geq 0 \Rightarrow \hspace{2ex} \left\{\pm
M^{\prime} \geq 0, \hspace{3ex}\left( E^{\prime} M - \frac{2}{3}
M^{\prime} E + \frac{T^{\prime}|E|^{5/3}}{3 \pi}\right) \geq 0,
\hspace{3ex} \pm T^{\prime} \leq 0 \right\}.    \label{asw}
   \ee
Relations (\ref{asy}) and (\ref{asw}) are the Hellaby and Lake
\cite{hel} no-shell-crossing conditions.

For values of $r$ at which $E(r) = - 1$, the condition $X \neq 0$
and finite implies that we must have
  $$
  \frac{R^{\prime}}{\sqrt{1 + E}} \neq 0
  $$
  and finite as $r \rightarrow r^{*}$ where
$E(r^{*}) = - 1$.  Again, differentiation of the solution for $R$
with respect to $r$ gives
\begin{quote}
$\lim_{r \rightarrow r^{*}} \frac{M^{\prime}}{\sqrt{1 + E}},
\hspace{2ex}  \lim_{r \rightarrow r^{*}}\frac{T^{\prime}}{\sqrt{1 + E}}
\hspace{2ex}  \lim_{r \rightarrow r^{*}}\frac{E^{\prime}}{\sqrt{1 + E}}$
\end{quote}
must be finite, and at least one must be non-zero.  For a non-zero
density at $r^{*}$, $\frac{M^{\prime}}{\sqrt{1 + E}}$ must be
non-zero.

More generally, $\mu = 0$ in {\bf s1} and {\bf s2} from their
definition.  In {\bf s3} and {\bf s4}, $\mu$ is only zero if
$M^{\prime} = 0$; otherwise $\mu$ is finite for all $r$ since
$R^{\prime}/M^{\prime} \neq 0$. For {\bf s5} it follows from above
that $R^{\prime}/M^{\prime} \neq 0$, and hence the density is
finite even where $E(r) = -1$.  For {\bf s6} the density vanishes
if $B = 0$ and $A \neq 0$, otherwise it is finite and positive if
and only if
 $$
  \lim_{\pm} \frac{B}{A} \geq \frac{1}{\pi}.
 $$

\ni We now consider the conditions for a centre $(R = 0)$ to exist
or be attached to a solution.  Only comoving centres are possible
and they may only join to solutions {\bf s1} to {\bf s5}.  A list
of the possibilities is given in table 1 \cite{hu1}.

\begin{table}
\begin{center}
\begin{tabular}{|l|cl|l|l|}\hline
Soln. &      & Behaviour of $E$, $M$ and $T$   & Kinematics &
Example\\ \hline\hline
         & & & & \\
         & (i)  & $T\rightarrow 0$                & $\Theta\equiv 0$ & \\
({\bf s1})     &      &                                 & &
$T=-r$\\
         & (ii)  & $\lim T^\prime$ finite, nonzero & $\mu\equiv 0$ &\\
         &      &                                 && \\
\hline
         & & & &\\
         & (i)  & $E\rightarrow 0$   & $\Theta\rightarrow 3\epsilon\tau^{-1}$
          & $E=r^2$\\
({\bf s2})     & (ii)  & $\lim (E^{-1/2}E^\prime)$ finite, nonzero
& &\\
         & (iii)  & $ET^\prime/E^\prime\rightarrow 0$  & $\mu\equiv 0$ &
         $T=0$\\
         &      &                                  & &\\
\hline
         & & & &\\
         & (i)  & $M\rightarrow 0$
         & $\Theta\rightarrow 2\epsilon\tau^{-1}$ & $M=r^3$\\
({\bf s3})     & (ii)  & $\lim (M^{-2/3}M^\prime)$ finite, nonzero
& & \\
         & (iii)
         & $MT^\prime/M^\prime\rightarrow 0$
          & $4\pi\mu\rightarrow\frac{2}{3}\tau^{-2}$ & $T=0$\\
         &      &     & & \\
\hline
         & & & & \\
         & (i)  & $E^{3/2}/M\rightarrow 0$, $M\rightarrow 0$
         & $\Theta\rightarrow 2\epsilon\tau^{-1}$ & $E=r^3$\\
         & (ii)  & $\lim (M^{-2/3}M^\prime)$ finite,
         nonzero  & & $M=r^3$\\
         & (iii)  & $\lim (MT^\prime/M^\prime)=\lim(M^{1/3}E^\prime/M^\prime)=0$
         & $4\pi\mu\rightarrow\frac{2}{3}\tau^{-2}$ & $T=0$\\
         &      &  & &\\
\cline{2-5}
         & & & &\\
         & (i)  & $E^{3/2}/M\rightarrow+\infty$, $E\rightarrow 0$
          & $\Theta\rightarrow 3\epsilon\tau^{-1}$ & $E=r^2$\\
({\bf s4})     & (ii)  & $\lim(E^{-1/2}E^\prime)$ finite, nonzero
& & $M=r^4$\\
         & (iii)  & $\lim(ET^\prime/E^\prime)$ & $4\pi\mu\rightarrow 0$
         & $T=0$\\
         &      & $=\lim\left[E^{-1/2}M^\prime E^{\prime -1}
         \log(E^{3/2}M^{-1})\right]=0$  & &\\
\cline{2-5}
         & & & &\\
         & (i)  & $E^{3/2}/M\rightarrow\alpha>0$, $M\rightarrow 0$
         & $\Theta\rightarrow 3\epsilon\alpha\sinh\eta/(\cosh\eta-1)^2$
         & $E=r^2$\\
         & (ii)  & $\lim(E^{-1}M^\prime)$ finite, nonzero & & $M=r^3$\\
         & (iii)  & $E^{-1}ME^\prime/M^\prime\rightarrow 2/3$
         and $MT^\prime/M^\prime\rightarrow 0$  & $4\pi\mu\rightarrow
         3\alpha^2/(\cosh\eta-1)^3$ & $T=0$ \\
         &      &  & &\\
\hline
         & & & &\\
         & (i)  & $|E|^{3/2}/M\rightarrow 0$,
          $M\rightarrow 0$ & $\Theta\rightarrow 2\epsilon\tau^{-1}$ &
          $E=-r^3$\\
         & (ii)  & $\lim(M^{-2/3}M^\prime)$ finite, nonzero & & $M=r^3$\\
         & (iii)  &
$\lim(MT^\prime/M^\prime)=\lim(M^{1/3}E^\prime/M^\prime)=0$
         & $4\pi\mu\rightarrow\frac{2}{3}\tau^{-2}$ & $T=0$\\
({\bf s5})     &      &  & &\\ \cline{2-5}
         & & & &\\
         & (i)  & $|E|^{3/2}/M\rightarrow\alpha>0$, $M\rightarrow 0$
     & $\Theta\rightarrow 3\epsilon\alpha\sin\eta/(1-\cos\eta)^2$ & $E=-r^2$\\
         & (ii)  & $\lim(E^{-1}M^\prime)$ finite, nonzero & & $M=r^3$\\
         & (iii)  & $E^{-1}ME^\prime/M^\prime\rightarrow 2/3$ and
         $MT^\prime/M^\prime\rightarrow 0$
         & $4\pi\mu\rightarrow 3\alpha^2/(1-\cos\eta)^3$ & $T=0$\\
         &      &    & &\\
\hline
\end{tabular}

~\\

 \caption{Central Behaviour}\label{centraltab}
\end{center}
\end{table}

In the table, results labelled (i) arise from the behaviour as $R
\rightarrow 0$; those labelled (ii) derive from the requirement
that $X \neq 0$, and (iii) are a result of the shear vanishing.
For each case, $M/R^{3} \rightarrow \frac{4}{3}\pi \mu$, which is
the Newtonian limit.  Simple examples for which the centre lies at
$r = 0$ are given in each case for illustration.  The expansion
rate at the centre,
 $$
 \Theta = 2\frac{\dot{R}}{R} + \frac{\dot{X}}{X},
  $$
  is listed for each solution. The central
behaviours listed for {\bf s4} and {\bf s5} generalise previous
results. An illustration of the calculations involved to derive
the results in the table is given in \cite{hu1}.

\section{Matching of Solutions}

In this section we reach the core of the paper. We examine
matching across comoving space-like surfaces between solutions
{\bf s1} to {\bf s6} to form composite models. The sign of
$R^{\prime}$ cannot change across these interfaces since $E \neq
-1$ on them - except on interfaces between {\bf s5} and {\bf s6}.
However note that $R^{\prime} \equiv 0$ in {\bf s6}. Solutions
{\bf s1} may not be matched to any others because they are
cosmological and it has $\dot{R} = 0$. Solution {\bf s2} does not
match to {\bf s5} since $(E \rightarrow 0$,$M \rightarrow 0)$
forces $R \rightarrow 0$ in {\bf s5}.  From the properties that
characterise {\bf s6} as an LTB model (see section 5.5), it
follows that it only matches to {\bf s5}.  There remain just five
physical types of junction.

\subsection{(a) Matching {\bf s2} to {\bf s4}}

\ni The {\bf s2} (interior) side of the junction is unconstrained
by the matching. Approaching the junction from {\bf s4}, $M
\rightarrow 0,$ $E > 0$ and $M^{\prime} > 0$ ($M^{\prime}$ is
continuous in {\bf s4}) since $M > 0$ in {\bf s4} and $M = 0$ in
{\bf s2}. Hence $R^{\prime} > 0$ and $R$ is increasing in the
direction {\bf s2} to {\bf s4}. Denote the value of $r$ at the
junction by $r^{*}$. Then as $r \rightarrow r^{*}$, $\eta
\rightarrow \infty$ because $M \rightarrow 0$ and $E
> 0$ in the exact solutions ({\bf s2}) and ({\bf s4}).  It follows that
  $$ X \rightarrow \frac{1}{\sqrt{1 +
E}}\left[\frac{M^{\prime}}{E}\log \left(\frac{E^{3/2}}{M} \right)
- T^{\prime} E^{1/2} + \frac{E^{\prime} \tau}{2E^{1/2}} \right]
  $$
  and hence that $X$ satisfies {\bf R2} if $\lim_{({\bf s4})}
T^{\prime}$,$\lim_{({\bf s4})} E^{\prime}$ and $\lim_{({\bf s4})}
M^{\prime} \ln M$ are finite and at least one is non-zero.

On the ({\bf s4}) side, the density reduces to
\be
4 \pi \mu \rightarrow \left[\tau^{2} \log \left(\frac{E^{2/3}
T^{\prime} \tau^{2}}{M} \right) - \frac{E^{3/2} T^{\prime}
\tau^{2}}{M^{\prime}} + \frac{E^{1/2} E^{\prime} \tau^{3}}{2
M^{\prime}} \right)^{-1},
 \ee
  which goes to zero as $ r
\rightarrow r^{*}$.  Note that $\mu = 0 $ in ({\bf s2}).

 Since on the ({\bf s4}) side $M \rightarrow 0$ as $r \rightarrow
r^{*}$ and $M^{\prime}M$ is finite, we must have $M^{\prime}
\rightarrow 0$ and hence
  $$ \frac{\dot{X}}{X} = \left\{
\begin{array}{ll}
                0 & \mbox{if $E^{\prime} \rightarrow 0$}, \\
                \eta \left[\tau - \frac{2 E T^{\prime}}{E^{\prime}} -
                \frac{2M^{\prime}}{E^{1/2} E^{\prime} \log M} \right]^{-1} &
                \mbox{otherwise}.
                \end{array}
 \right.
 $$
 On the ({\bf s2}) side
 $$
 \frac{\dot{X}}{X} \rightarrow \frac{\epsilon}{\tau -
 2ET^{\prime}/E^{\prime}},
 $$
 and on both sides $\dot{R}/{R} \rightarrow \epsilon/\tau$.  From
 this it follows that the shear remains finite on both sides of
 the junction.
\subsection{(b) Matching {\bf s3} to {\bf s4}}

\ni The {\bf s3} side is unconstrained by the matching.
Approaching the junction from within the {\bf s4} region, $E
\rightarrow 0$ and $M > 0$. Since $E > 0$ in {\bf s4}, we must
have $E^{\prime} > 0$ in a neighbourhood of the junction in the
{\bf s4} region.  Also, from the explicit solution for {\bf s4} we
have
  $$
  \sinh \eta - \eta = \tau E^{3/2} M^{-1} \rightarrow 0,
  $$
because $E \rightarrow 0$ and $M > 0$.  Thus $\eta \approx (6
\tau/M)^{1/3} E^{1/2}$ near the junction on the {\bf s4} side and
$E^{\prime} > 0$, since $E > 0$ in the ({\bf s4}) region. Hence
$R$ must increase in the direction ({\bf s3}) to ({\bf s4}).
 On the ({\bf s4}) side,
 \be
 X \rightarrow M^{\prime} \left(\frac{\tau^{2}}{6 M^{2}}
 \right)^{1/3} - T^{\prime} \left(\frac{4M}{3 \tau} \right)^{1/3}
 + \frac{E^{\prime}}{40}\left(\frac{(6\tau)^{4}}{M} \right)^{1/3},
 \label{s34x}
 \ee
 and so $X$ is finite and non-zero provided
 \begin{quote}
$\lim_{({\bf s4})}M^{\prime}$, $\lim_{({\bf s4})}T^{\prime}$ and
$\lim_{({\bf s4})}E^{\prime}$ are finite and at least one is
non-zero.
 \end{quote}
 On both sides of the interface, the density reduces to
 \be
 4 \pi \mu \rightarrow  \left\{ \begin{array}{ll}
                0 & \mbox{if $M^{\prime} \rightarrow 0$}, \\
                \left[\frac{3}{2}\tau^{2} - \frac{3 M T^{\prime}
                \tau}{M^{\prime}} -
                \frac{(6 \tau)^{8/3} M^{1/3}E^{\prime}}{160 M^{\prime}}
                \right]^{-1} &
                \mbox{otherwise},
                \end{array}
 \right. \label{s34m}
 \ee
 and on both sides,
 \bea
 \frac{\dot{X}}{X} & \rightarrow & \frac{\epsilon \left(T^{\prime} +
 \frac{M^{\prime} \tau}{M} + \left(\frac{243 \tau^{5}}{250 M^{2}}
\right)^{1/3} E^{\prime} \right)}{\left(\frac{3 M^{\prime}
\tau^{2}}{2M} - 3T^{\prime} \tau + \frac{E^{\prime}}{10
M^{2/3}}\left(\frac{9 \tau^{2}}{2}
 \right)^{4/3} \right)},  \label{s34y} \\
 \frac{\dot{R}}{R} & \rightarrow & \frac{2 \epsilon}{3 \tau}.
 \label{s34r}
 \eea
 Therefore $\mu$ and $\frac{\dot{X}}{X}$ and the shear are regular up to
 the junction.

\subsection{(c) Matching {\bf s3} to {\bf s5}}

This junction is similar to (b). On approaching the junction in
({\bf s5}), $\eta \approx (6 \tau M)^{1/3} |E|^{1/2} \rightarrow
0.$ All the results in section 5.2 for the kinematics and metric
components follow with ({\bf s4}) replaced by ({\bf s5}).

In this case $R$ must increase in the opposite sense to that in
section 5.2, i.e., here $R$ must increase in the direction ({\bf
s5}) to ({\bf s3}), as we will now show.  Working in ({\bf s5}),
at the junction $E = 0$, and since $E < 0 $, it follows that
$E^{\prime} < 0$ in a neighbourhood of the junction.  Also, near
the junction the conditions (\ref{asw}) hold, so $R^{\prime}$ and
$M^{\prime}$ have the same sign and since $M > 0$ and $E$ may be
as small as we please, $R^{\prime}$ and $E^{\prime}$ have the same
sign.  Since $E^{\prime} < 0$, we have $R^{\prime} < 0$, i.e., $R$
increases in the direction from ({\bf s5}) to ({\bf s3}).  It is
interesting to note that, since $\dot{R}/R \rightarrow \frac{2}{3}
\epsilon \tau^{-1} > 0$ at a junction between ({\bf s3}) and ({\bf
s5}), the continuity of $\dot{R}/R$ forces the existence of a
finite region in ({\bf s5}), adjoining the junction, where the
azimuthal expansion rate $\dot{R}/R$ is positive even though all
points in ({\bf s5}) eventually satisfy $\dot{R} < 0$.

\subsection{(d) Matching {\bf s4} to {\bf s5}}

Both sides are constrained by $E \rightarrow 0$ with $M > 0$ and
the result is obtained by combining results from ({\bf b}) and
({\bf c}). In this case $R$ must increase in the direction ({\bf
s5}) to ({\bf s4}) by a similar argument to that given in ({\bf
b}).

\subsection{(e) Matching {\bf s5} to {\bf s6}}

For solutions ({\bf s6}), $R^{\prime} = 0$, and so they may only
be matched, across a comoving surface, to solutions of the type
({\bf s5}), because $R^{\prime} = 0$, $X \neq 0$ requires $E =
-1$. This motivates a characterisation of ({\bf s6}) within the
family of LTB solutions by the conditions $M^{\prime} = T^{\prime}
= 0$, $M > 0$ and $E = -1$.

At this junction the ({\bf s6}) side is unconstrained by the
matching.  The observer area distance $R$ may increase in either
direction on approaching the junction from ({\bf s5}).  Both
metrics are regular on approach to the junction and it is less
restrictive than the other four. \\

When the conditions for matching and for a centre are combined,
four classes emerge.

\section{Models by Class}

\subsection{Open models with one centre}

By noting the sense in which $R$ must increase at the interfaces
({\bf a}) to ({\bf e}) above, the only possible composite models
are:
\[\begin{array}{ll}
{\cal O}{\rm({\bf s1})}^+,    & {\cal O}{\rm({\bf s2})}^+{\rm({\bf
s4})}^+,\\ {\cal O}{\rm({\bf s2})}^+,    & {\cal O}{\rm({\bf
s3})}^+{\rm({\bf s4})}^+,\\ {\cal O}{\rm({\bf s3})}^+,    & {\cal
O}{\rm({\bf s5})}^+{\cal S}{\rm ({\bf s5})}^+ {\rm({\bf s3})}^+,\\
{\cal O}{\rm({\bf s4})}^+, & {\cal O}{\rm({\bf s5})}^+{\cal S}{\rm
({\bf s5})}^+ {\rm({\bf s4})}^+,\\ {\cal O}{\rm({\bf s5})}^+{\cal
S},~~~~~~~~~~ & {\cal O}{\rm({\bf s5})}^+{\cal S}{\rm ({\bf
s5})}^+{\rm({\bf s3})}^+{\rm({\bf s4})}^+,
\end{array}\]
where ${\cal O}$ denotes a centre, and a superscript $+$ ($-$)
implies that $R$ increases (decreases) from left to right. Here
${\cal S}$ is {\it any} combination of ${\rm({\bf s5})}^-$,
${\rm({\bf s5})}^+$ and ${\rm({\bf s6})}$. Note that open models
can be constructed from collapsing solutions [e.g. ${\cal
O}{\rm({\bf s5})}^+{\rm({\bf s6})}$]. Papapetrou \cite{pap}
discussed a particular example of ${\cal O}{\rm({\bf
s5})}^+{\rm({\bf s3})}$.

In the above construction, we have noted from (\ref{fe2}) that on
$t=$const, $d\chi=|dR|/\sqrt{1+E}$, where $\chi$ is radial proper
distance. Hence by (\ref{15}), if $E>\alpha>-1$ for all
$\chi>\beta$ ($\alpha,\beta$ constants) then:
\be\chi\rightarrow\infty~~{\rm forces}~~R\rightarrow\infty~~~~~
{\rm if}~~\frac{dR}{d\chi}>0,\label{asymp1}\ee
\be{\rm there~is~a~finite~value~of~}\chi>\beta~{\rm~for~which~}
R=0,~~~~~{\rm if}~\frac{dR}{d\chi}<0.\label{asymp2}\ee However, if
$E\rightarrow-1$ as $\chi\rightarrow\infty$, then neither of
(\ref{asymp1}),({\ref{asymp2}) are necessary.\\

An example of ${\cal O}{\rm({\bf s5})}^+{\rm({\bf s5})}^-$ in the
class of open models with one centre is
  \bea
  E & = &
\left\{\begin{array}{lll} -\frac{\sin^2r}{\sin^2r_0}\left[1-{\rm
e}^{-2r_0}\right]&{\rm for}&0<r<r_0,\\ -1+{\rm e}^{-2r}&{\rm
for}&r>r_0,
\end{array}\right.~~~ \\
M & = & \left\{\begin{array}{lll}
\frac{\sin^3r}{\sin^3r_0}\left[M_\infty+{\rm e}^{-r_0}\right]&{\rm
for}& 0<r<r_0,\\ M_\infty+{\rm e}^{-r}&{\rm for}&r>r_0,
\end{array}\right. \\
T & = & 0,~~~M_\infty>0,~~~\pi<r_0<2\pi,
\label{example1}
\eea
and
\bea
E & = & \left\{\begin{array}{lll}
-\frac{r^2}{r_0^2}\left[1-{\rm e}^{-2r_0}\right]&{\rm for}&0<r<r_0,\\
-1+{\rm e}^{-2r}&{\rm for}&r>r_0,
\end{array}\right.~~~ \\
M & = & \left\{\begin{array}{lll}
\frac{r^3}{r_0^3}\left[M_\infty-{\rm e}^{-r_0}\right]&{\rm for}&
0<r<r_0,\\
M_\infty-{\rm e}^{-r}&{\rm for}&r>r_0,
\end{array}\right. \\
T & = & 0,~~~0<M_\infty<2/3,~~~r_0>0, \label{example2}\eea is an
example of ${\cal O}{\rm({\bf s5})}^+$. In each of
(\ref{example1}) and (\ref{example2}), $R\rightarrow$const$>0$ as
$\chi\rightarrow\infty$. There are no spherically symmetric dust
models with $R\rightarrow 0$ as $\chi\rightarrow\infty$ [by
(\ref{asymp1}) and since, for the exact solutions, $R\rightarrow
0$ requires $E\rightarrow 0$].

\subsection{Open models with no centre}

\ni By (\ref{asymp2}), to avoid a zero in $R$, a model with no
centre must either be composed entirely of ({\bf s6}), or it must
contain a section of ({\bf s5}), in order to allow (at least one)
minimum in $R$. Then the possible matchings are evident:
\[\left.\begin{array}{r}
{\cal S}\\ {\rm({\bf s3})}^-{\rm({\bf s5)}}^-\\ {\rm({\bf
s4})}^-{\rm({\bf s5})}^-\\ {\rm({\bf s4})}^-{\rm({\bf
s3})}^-{\rm({\bf s5})}^-\end{array}\right\} {\cal
S}\left\{\begin{array}{l} {\cal S}\\ {\rm({\bf s5})}^+{\rm({\bf
s3})}^+\\ {\rm({\bf s5})}^+{\rm({\bf s4})}^+\\ {\rm({\bf
s5})}^+{\rm({\bf s3})}^+{\rm({\bf s4})}^+\end{array}\right.\]
Examples and a detailed analysis of such models are provided in
\cite{h87}. In these models, due to the presence of collapsing
solutions ({\bf s5}),({\bf s6}), a centre does eventually form,
but gravitational collapse will violate the regularity conditions
in any case.

\subsection{Closed models with two centres}

\ni  These models must contain a region of ({\bf s5}), since there
must be (at least one) turning point in $R$. The models cannot
contain a region of ({\bf s2}), ({\bf s4}) or ({\bf s2})({\bf
s4}), since the region would either contain a centre and match to
another solution, or would match to other solutions on both sides.
Hence $E$ would vanish on both sides, and since $E>0$ throughout
the domains of ({\bf s2}) and ({\bf s4}), $E^\prime$ could not
have the same sign throughout, contrary to (\ref{asy}) [with
(\ref{15})]. There can be no ({\bf s1}) region in the closed
model, since it does not match to any other solution. There can be
no ({\bf s3}) region in the model either, since $R$ must increase
in the direction ({\bf s5})$\to$({\bf s3}). Hence if ({\bf s3})
contains a centre, it cannot match to ({\bf s5}). Conversely, if
({\bf s3}) does not contain a centre, it cannot match to ({\bf
s5}) on both sides, leaving the model open. This leaves just ({\bf
s5}) and ({\bf s6}) to construct these models, and the
possibilities are:
\[{\cal O}{\rm({\bf s5})}^+{\cal S}{\rm({\bf s5})}^-{\cal O}\]

\subsection{Closed models with no centre}

Consider an SS dust model which has $R>0$ in some range $0\leq
r\leq d$ (and at some $t$). This final possibility of composite
models is obtained by identifying (matching) the surfaces $r=0$
and $r=d$. Since $\Delta R=0$, the model must be everywhere ({\bf
s6}) or else it must contain a region of ({\bf s5}) [otherwise
${\rm sign}(R^\prime)$ is constant in $0\leq r\leq d$, which
forces $R(0)\neq R(d)$]. No regions composed from the solutions
({\bf s1})-({\bf s4}) may be present, since they would be forced
to match to ({\bf s5}) on both sides. This would force $R^\prime$
to change sign in the region (since $R$ must increase away from
({\bf s5}) into these solutions) and this is not possible, by
(\ref{15}). Hence the models may only be constructed from ({\bf
s5}) and ({\bf s6}), with the possibilities: \\
\[ {\cal I}{\cal S}{\cal I} \], \\
where ${\cal I}$ denotes the surfaces which are identified (at
which the standard matching conditions must be satisfied, as we
have described). The spatial sections of these models have the
topology of a 3-torus. An example is provided by \bea
 E & = & \left\{\begin{array}{ll}
ar^2-1      & {\rm for}~0<r<\frac{1}{4}d, \\
&\\
a(r-\frac{1}{2}d)^2-1 & {\rm for}~\frac{1}{4}d<r<\frac{3}{4}d, \\
&\\
a(r-d)^2-1 & {\rm for}~\frac{3}{4}d<r<d,
\end{array}\right.~~~ \\
\vspace{2ex}
M & = &\left\{
\begin{array}{ll}
b+cr^2 & {\rm for}~0<r<\frac{1}{4}d, \\ &\\
b+\frac{1}{8}cd^2-c(r-\frac{1}{2}d)^2 & {\rm for}~
\frac{1}{4}d<r<\frac{3}{4}d, \\ &\\ b+c(r-d)^2 & {\rm
for}~\frac{3}{4}d<r<d,\end{array}\right.  \\
 T& = & 0,~~~a\left(2b+\frac{1}{4}cd^2\right)<\frac{4}{3}c,
 \eea
where $a,..,d$ are positive constants. Note that a closed model
with no centre
 cannot be constructed from the homogeneous
(Friedmann-Lema\^{\i}tre-Robertson-Walker) subclass of LTB (since
the elliptic homogeneous solution has only one point with $E=-1$,
at which $R$ is maximum).
\\

There are no further possible classes or composite models.  There
can be a number of different topologies constructed from these
models but that involves different questions from those tackled
here.   Examples of models of types described in sections {\bf
6.1} to {\bf 6.3} are given in previous literature (see especially
\cite{h87}).

\section{A Black Hole in an Expanding Universe}

In this section we present an example which serves to illustrate
some points of significance in using matched LTB models in
cosmology. First that combinations of different LTB models can
provide realistic exact models of cosmologically interesting
phenomena; second that the richness of the models is often
overlooked; thirdly that even in simple cases, there are subtle
issues that need to be watched (for instance an apparently useful
model may be flawed) and fourthly the illustration itself has some
intrinsic interest as a model of the formation of a black hole in
an Einstein-de Sitter space-time which is different from the well
known Oppenheimer-Snyder case.

In his treatise on inhomogeneous cosmological models \cite{kr1}
Krasi\'{n}ski gives the history of attempts to describe the
formation of black holes in an expanding universe.  The first
successful attempt was by Barnes \cite{bar}, who proved sufficient
conditions for a black hole to form and he studied several
examples of collapse in LTB coordinates.  Subsequent work on
similar lines was done by Demia\'{n}ski and Lasota \cite{dem} and
Polnarev \cite{pol}.  In 1978, Papapetrou published a model of a
collapsing region in an Einstein-de Sitter spacetime. In his model
the formation of the apparent horizon and the central singularity
can easily be followed.  In this model and in all others known to
us, the singularity goes on accreting matter for an infinite LTB
time, which limits their value in exact cosmology. In fact,
although it is not widely known, this must happen irrespective of
the precise nature of the collapsing part, unless it contains a
vacuum region.  The result follows from the matching conditions.
If an {\bf s5} region is matched to either an {\bf s3} or {\bf s4}
one then, as we have seen, $E$ must vanish at the boundary. It
follows that when $\eta = 2 \pi $, $ \tau$ must be infinite on the
boundary of the elliptic region ({\bf s5}). Therefore, if the
matter inside the elliptic region extends to its boundary then
collapse to a singularity will take an infinite time to complete.
We overcome this difficulty by arranging for the matter to extend
only to some value of $r$ within the elliptic ({\bf s5}) region.

Although the end result in our model is an Einstein-Strauss
vacuole the dynamics of the formation of the hole is different. In
Oppenheimer-Snyder collapse, the mass gets trapped at the boundary
first and at the centre last.  In our case, the reverse happens
and the horizon begins to form at the centre and spreads outwards,
and the singularity is covered at all times.

\subsection{The Model}

To define the model we choose the following arbitrary functions:
\bea
 T(r) & = & 0  \hspace{6ex} \forall r, \\
 E(r) & = & \left\{ \begin{array}{ll}
                        - \beta\left(\frac{r}{r_{0}} \right)^{2}
                        \left(1 - \frac{r}{r_{0}} \right)^{2} &
                        0 \leq r \leq r_{0}, \\
                0 & r_{0} < r,
                \end{array} \right. \\
 M(r) & = & \left\{ \begin{array}{ll}
                       \frac{1}{2} \alpha r^{3}     &     r_{0} < r, \\
         \frac{1}{2}\alpha r_{1}^{3} &     r_{1} \leq r \leq r_{0}, \\
          \frac{1}{2}\alpha r^{3} \left(\frac{r_{1}}{r_{0}}\right)^{3}
                                        &      r_{0} < r.
                        \end{array} \right. .
\eea The four arbitrary constants in the model are limited as
follows:
\begin{enumerate}
\item $ \alpha > 0 $
\item $ 16 > \beta > 0 $
\item $ r_{1} > \frac{1}{2} r_{0} $
\end{enumerate}
The following properties of the model follow easily from well
known results:
\begin{enumerate}
\item[a.] From  a formula of Barnes \cite{bar},
$$ \frac{R^{\prime}}{R} = \left(\frac{M^{\prime}}{M} -
\frac{E^{\prime}}{E}\right) - \left( \frac{M^{\prime}}{M} -
\frac{3 E^{\prime}}{2 E} \right)\frac{t \dot{R}}{R},
 $$
  and the
above, plus the appropriate exact solutions for $R$, it follows
that $R^{\prime}> 0 $, as required to avoid a singularity in the
metric.
\item[b.] From condition (1) and $R^{\prime} > 0$, the density is
positive.
\item[c.] From (2) it follows that $(1 + E) > 0$,  which ensures that
$r$ remains spacelike.
\item[d.]
The junction conditions are satisfied for the (combined) metric, and it
is non-singular except at $\eta = 0$ or $2 \pi$.
\end{enumerate}

For $ r > r_{0}$, the model is an Einstein-de Sitter universe with
density $\rho = (6 \pi t^{2})^{-1}$.  For $0 \leq r \leq r_{0}$,
it represents an elliptic region which first expands and then
collapses.  In the region $ r_{1} < r \leq r_{0}$, it is vacuum,
and for $ 0 \leq r \leq r_{1}$, it contains dust.

\subsection{Collapse}

Here we will discuss the dynamics of the elliptic region in some
detail.  It starts with a big bang at $t = 0$ and all shells
expand until they reach their maximum surface area which occurs at
$\eta = \pi$, i.e. at times given by
\be
t = \frac{\pi M}{(- E)^{3/2}} \ee which is a monotonically
increasing function of $r$.  This means that shells with larger
values of $r$ reach their maximum later than those with smaller
$r$.  The shell bounding the matter, $r = r_{1}$, reaches its
maximum at time
\be
 t = \frac{\pi \alpha r_{0}^{3}}{2[\beta^{1/2}(1 -
 \frac{r_{1}}{r_{0}})]^{3}}.
\ee

After reaching their maximum surface area, the shells collapse to
a singularity at $R = 0$, which occurs when $\eta = 2 \pi$, i.e.
at time
\be
t = \frac{\pi 2 M}{(- E)^{3/2}},
 \ee
  which depends on $r$.  The
point $r = 0$ is exceptional in that $R$ vanishes for all $t$.
However the behaviour of the density shows that after the
universal singularity at $t = 0$, the point is non-singular until
$\eta = 2 \pi$.  In LTB time, the singularity at $r = 0$ begins to
form at time  $t(0) = t_{0}$, where, taking limits,
\be
t_{0} = \frac{\pi \alpha r_{0}^{3}}{\beta^{3/2}}. \ee Note that
the $t_{0}$ here is twice that found by Papapetrou \cite{pap}.
The collapse process continues with the shells labelled $r$
becoming singular at time
\be
t = t_{0} \left(1 - \frac{r}{r_{0}} \right)^{-3}.  \label{19}
\ee
The last shell at $r = r_{1} $ becomes singular at $t_{1}$ given
by
\be
t_{1} = t_{0}\left(1 - \frac{r_{1}}{r_{0}} \right)^{-3},
 \ee
 after
which the collapse is complete.

\ni The mass in the singularity at any time $t_{0} \leq t \leq
t_{1}$ is given by
\be
M = \frac{1}{2} \alpha r_{0}^{3} \left[1 -
\left(\frac{t_{0}}{t}\right)^{1/3}\right]^{3}
 \ee
  Therefore at the
beginning of the collapse, $M(t_{0}) = 0$ and at the end,
$M(t_{1}) = \frac{1}{2}\alpha r_{1}^{3}$.  Given the size of a
spherically symmetric tophat region and the mass enclosed, we
could determine $\alpha$.

After time $t_{1}$ the solution is only defined for $r > r_{1}$,
i.e., in the exterior vacuum and Einstein-de Sitter regions.  The
vacuum represents the Schwarzschild region in comoving
coordinates. \footnote{Singularities continue to form for $r >
r_{1}$ at times $$ t = t_{0}\left(\frac{r_{1}}{r_{0}} \right)^{3}
\left( 1 - \frac{r}{r_{0}} \right)^{-3}, $$ and are to be
interpreted as the arrival of successive shells of {\em test
particles} which label the coordinates of the vacuum in this
gauge.  This process continues until the shell $r = r_{0}$ arrives
at time $t = \infty$.}

For $t > t_{1}$, the solution is equivalent to an Einstein-Strauss
vacuole in an Einstein-de Sitter universe \cite{ein}.

\subsection{Horizons}

As usual in LTB spacetimes, we use the apparent horizon as
diagnostic for the existence of an event horizon and therefore a
black hole. The apparent horizon is given by
\cite{bar}\footnote{This is the apparent horizon bounding trapped
surfaces associated with the collapse, which begins at $\eta =
\pi$.  There is also an apparent horizon related to the initial
expansion which all particles must cross during their expansion
phase from the big bang {\em white hole}.  Gautreau and Cohen
\cite{gau} call this a boundary of expelled surfaces.}
\be
R(r,t) = 2M(r),  \hspace{6ex} \dot{R} < 0 . \label{23}
 \ee
  This formula can be put in the alternative forms
\be
 \dot{R} = - (1 = E)^{1/2}
\ee
and
\be
\sin (\eta/2) = (- E)^{1/2}, \hspace{6ex} \pi < \eta \leq 2 \pi.
\label{25}
 \ee
  To simplify the notation, we define $w :=
\frac{r}{r_{0}}$.  Then (\ref{25}) gives
\be
\sin(\eta/2) = \beta^{1/2} w ( 1 - w), \hspace{6ex} \pi < \eta
\leq 2 \pi,
 \label{27}
\ee and from the definition of $\eta$ in solution ({\bf s5}) and
the definitions of the arbitrary functions,
 \bea
t_{AH}(r)& = &  \frac{t_{0}(\eta - \sin \eta )}{2 \pi(1 - w)^{3}},
\hspace{6ex} 0 \leq w \leq w_{1}, \label{ah1} \\
 t_{AH}(r) & = & \left( \frac{t_{0}(\eta - \sin \eta )}{2 \pi(1 - w)^{3}}
\right)  \left( \frac{w_{1}}{w}\right), \hspace{3ex} w_{1} \leq w
\leq 1, \label{ah2} \eea where $w_{1} := r_{1}/r_{0}$ and $\eta$
is determined by (\ref{27}).  Equations (\ref{ah1}) and
(\ref{ah2}) together give the equation of the apparent horizon.
The areal radius of the apparent horizon at coordinate $r$ is
given by
 \bea
 R_{AH} & = & \alpha r^{3}, \hspace{6ex} 0 \leq r \leq r_{1}, \\
 R_{AH} & = & \alpha r_{1}^{3}, \hspace{5ex} r_{1} \leq r \leq
 r_{0},
\eea
where we have used the definition of $M$ for the appropriate range
of $r$ and equation (\ref{23}).

From (\ref{27}) if $w = 0$  then $\eta = 2 \pi$ and so (\ref{ah1})
gives $t_{AH} = t_{0}$.  Thus at $r = 0$ the singularity and the
apparent horizon form together.  When this happens, the
singularity may be naked \cite{ear,chr,dwi}, but we will not
discuss that here.  We will concentrate instead on the formation
of the black hole and assume that $0 < w \leq 1$.

We denote the time at which the singularity forms at $r$ by
$t_{s}(r)$, which is given by (\ref{19}), i.e.,
\be
t_{s}(r) = t_{0}( 1 - w)^{-3}.
 \ee
 Then for $0 \leq r \leq r_{1}$,
i.e., inside the collapsing dust sphere , we obtain from
(\ref{ah1})
\be
t_{s}(r) - t_{AH}(r) = \frac{t_{0}(2 \pi - \eta + \sin \eta)}{2
\pi (1 - w)^{3}}, \label{32}
 \ee
 and for $r_{1} \leq r \leq
r_{0}$, i.e., in the surrounding vacuum,
\be
t_{s}(r) - t_{AH}(r) = \frac{t_{0}(2 \pi - \eta + \sin \eta)}{2
\pi (1 - w)^{3}} \left(\frac{w_{1}}{w} \right)^{3}, \label{33}
\ee
where $\eta$ is given by equation (\ref{27}).  Given the range of
$\eta$, $t_{s} - t_{AH} > 0$ always.  So the apparent horizon
forms first and the singularity is not naked.  This agrees with a
result of Joshi \cite{jos}.  The horizons starts to form at the
centre and spreads outwards with time.  This is different from the
behaviour of the Oppenheimer-Snyder solution for which the
boundary and the mass get trapped first and the centre last.

In the limiting case where $w \rightarrow 1$, the inner collapsing
sphere extends to the Einstein-de Sitter exterior, so that there
is no vacuum region.  This is the case considered by Papapetrou
and analysis of the limits confirms his result.  Both $t_{s}$ and
$t_{AH}$ tend to $\infty$.  However the difference between them
remains finite and $t_{s}$ remains greater than $t_{AH}$.

\section{Conclusion}

\ni Here a set of spherically symmetric inhomogeneous dust models
has been provided which can be used to construct cosmological
solutions to Einstein's field equations for a range of
astrophysical situations including voids and black holes.  A
section of the book by Krazi\'nski \cite{kr1} is devoted to
applications of these models in cosmology. In setting out the
models new results have been obtained which fill gaps in the
literature. In particular we have demonstrated the matching of the
Kantowski-Sachs solutions and derived a new solution where the
spatial sections have a torus topology. Also the particular
regularity conditions we use lead to a restricted but, we would
argue, cosmologically more useful set of matched solutions. The
categorisation we provide of the allowed cases with centres is
useful for the construction of models.

The value of these matched solutions is demonstrated in the final
example of black hole formation which illustrates very clearly why
it is important to consider details when using inhomogeneous
models.  We have shown that using matched LTB solutions a model
can be constructed to describe collapse to a black hole in an
Einstein de Sitter background but it must contain a vacuum region
if the singularity is not to continue to accrete matter for an
infinite LTB time. \\

\ni {\bf Acknowledgements}

The authors thank W. B. Bonnor for suggesting the model in section
7 and for enlightening comments and discussions on this work and
Roy Maartens for helpful comments.

 \end{document}